\documentclass[12pt,a4paper,preprint,notitlepage,superscriptaddress]{revtex4-2}

\usepackage{amsmath,amssymb}
\usepackage{graphicx}
\usepackage{xcolor}
\ifdefined\CleanVersion
  \newcommand{\rev}[1]{\textcolor{black}{#1}}
  \newenvironment{revblock}{\begingroup\color{black}}{\endgroup}
\else
  \newcommand{\rev}[1]{\textcolor{red}{#1}}
  \newenvironment{revblock}{\begingroup\color{red}}{\endgroup}
\fi

\ifdefined\CleanVersion
  \newcommand{\chg}[1]{\textcolor{black}{#1}}
\else
  \newcommand{\chg}[1]{\textcolor{red}{#1}}
\fi

\usepackage{flafter}
\usepackage{placeins}
\usepackage{float}
\usepackage{hyperref}
\usepackage{orcidlink}
\usepackage[margin=1in]{geometry}
\usepackage{xurl}
\renewcommand{\doi}[1]{\href{https://doi.org/#1}{\nolinkurl{doi:#1}}}

\newcommand{\ManuscriptAffiliation}{\chg{Department of Matter Sciences, }Echahid Cheikh Larbi Tebessi University, Tebessa \chg{12002}, Algeria}

\newcommand{\FundingStatement}{}

\makeatletter
\let\GDO@abstract@produce\frontmatter@abstract@produce
\def\frontmatter@abstract@produce{%
  \begingroup
  \@booleanfalse\preprintsty@sw
  \GDO@abstract@produce
  \endgroup
}
\makeatother
\hypersetup{
  hidelinks,
  pdftitle={The Generalized Dirac Oscillator in Doubly Special Relativity: A Complexified Morse Interaction},
  pdfauthor={Abdelmalek Boumali}
}

\begin{document}

\title{The Generalized Dirac Oscillator in Doubly Special Relativity: A Complexified Morse Interaction}

\author{Abdelmalek Boumali\,\orcidlink{0000-0003-2552-0427}}
\email{boumali.abdelmalek@gmail.com}
\affiliation{\ManuscriptAffiliation}

\begin{abstract}
\rev{We \chg{study} the one-dimensional generalized Dirac oscillator \chg{with a complex Morse interaction, $f(x)=D-qe^{-\alpha x}$, in the Magueijo--Smolin} (\chg{MS}) \chg{and Amelino-Camelia (AC)} realizations of doubly special relativity (DSR). The spatial problem factorizes into supersymmetric \chg{partners}. For $D>0$ and \chg{$\mathrm{Re}\,q>0$,} the \chg{positive} metric $\eta=\exp(-\theta p/\hbar)$, \chg{$\theta=2\arg q/\alpha$}, \chg{defines, on analytic bound states, a} similarity transformation \chg{$\rho=\eta^{1/2}$} to a Hermitian Morse representative. \chg{Direct Kummer analysis on} the \chg{complex ray gives the real spectrum $\epsilon_n=D^2-(D-n\hbar\alpha)^2$}, independent of the phase of $q$, \chg{with $0\le n<D/(\hbar\alpha)$ following} from square integrability on the whole line. We \chg{relate} the Morse parameters to the DSR scale $k$. \chg{Both DSR} equations \chg{preserve the} spatial spectrum but \chg{deform energy} reconstruction and spinor amplitudes. \chg{On} admissible branches, \chg{$E^{\rm MS}_{n,\sigma}=\sigma\Omega_n-m^2/k$ and $E^{\rm AC}_{n,\sigma}=\sigma\Omega_n+\epsilon_n/(2k)$ to first order, with $\Omega_n=\sqrt{m^2+\epsilon_n}$}. \chg{On nonzero SUSY doublets,} the AC \chg{generalized-eigenvalue weight is positive exactly for $\epsilon_n<4k^2$. At this boundary} the \chg{positive} branch diverges \chg{while} the negative branch \chg{remains finite, yielding} a \chg{Morse excitation threshold}. \chg{Quantitative} use requires $|E|/k\ll1$. \chg{The model provides} an \chg{exactly solvable testbed linking} pseudo-Hermiticity, \chg{supersymmetry}, and DSR \chg{kinematics}.}
\end{abstract}

\maketitle
\raggedbottom

\section{Introduction}

Exactly solvable relativistic bound-state models are useful laboratories for separating genuinely relativistic effects from those produced by a specific interaction.  The Dirac oscillator (DO) is a canonical example.  Early linear-trajectory constructions were discussed by It\^{o}, Mori, and Carri\`ere \cite{Ito1967}, and the model was later formulated in its standard form by Moshinsky and Szczepaniak \cite{Moshinsky1989}.  Its algebraic structure has led to connections with quantum-optical models, including the Jaynes--Cummings interaction \cite{Bermudez2007}, as well as \rev{\chg{proposals for} photonic and other \chg{analogues}} \cite{Longhi2010,Sadurni2011}.

The generalized Dirac oscillator (GDO) replaces the linear oscillator coupling by a general function $f(x)$.  In $(1+1)$ dimensions the resulting Dirac equation reduces to a pair of intertwined Schr\"odinger-type problems.  This factorization places the GDO naturally within supersymmetric quantum mechanics (SUSY QM), where shape-invariant choices of $f(x)$ generate exactly solvable relativistic models \cite{Cooper1995,Gendenshtein1983,Junker2020}.  The Morse family is particularly useful because it combines analytic solvability with an intrinsically finite bound-state ladder.

A second structural ingredient appears when $f(x)$ is complex.  The Dirac Hamiltonian is then non-Hermitian in the standard Hilbert-space inner product, but spectral reality may survive.  Two concepts are relevant and should be distinguished.  A Hamiltonian may possess an unbroken antilinear $\mathcal{PT}$ symmetry \cite{Bender1998,Bender2007}, or it may be pseudo-Hermitian with respect to a positive metric operator \cite{Mostafazadeh2002,Mostafazadeh2010}.  These notions overlap in important classes of models but they are not identical.  In the present paper, the specific complex Morse interaction is treated by an explicit imaginary-translation metric, following the pseudo-Hermitian mechanism used for complex potentials and generalized Dirac oscillators \cite{Ahmed2001,Dutta2013}.  \rev{For this Morse family the same discrete quantization also follows directly on a complex ray of the exponential variable introduced in Sec.~IV\,B; conventional real-axis parity symmetry is therefore not part of the spectral mechanism.}  This distinction is important because the exponential Morse interaction is not, in general, invariant under the standard transformation $x\mapsto -x$.  \rev{Unlike the historical Morse problem, which is posed for a positive radial coordinate \cite{Morse1929}, the present model is a genuinely one-dimensional Dirac problem on the whole line $x\in\mathbb{R}$; the resulting lack of reflection invariance, the admissible range of the parameters, and their physical scales are made explicit in Sec.~IV\,A.}

The third ingredient is doubly special relativity (DSR), in which the relativistic kinematics contains a second observer-independent scale $k$ in addition to the speed of light.  Different nonlinear momentum-space realizations lead to different modified dispersion relations and wave equations \cite{AmelinoCamelia2002a,MagueijoSmolin2002,KowalskiGlikman2005,AmelinoCamelia2010,AmelinoCamelia2013}.  Relativistic oscillators provide a tractable setting in which these differences can be analyzed analytically.  Recent work has compared the Magueijo--Smolin (MS) and \chg{Amelino-Camelia} (AC) realizations for Dirac and Klein--Gordon oscillators and related thermal systems \cite{JafariBoumali2025PLB,BoumaliJafari2026CTP,BoumaliJafari2025EPJC,BoumaliJafari2026JLTP}.

The present work develops a modular spectral construction in which the non-Hermitian spatial problem is solved first and the DSR kinematics is imposed afterward through an energy-dependent algebraic map.  This viewpoint leads to several concrete results that sharpen the physical content of the model.  \rev{\chg{First, building on the complex-Morse generalized Dirac oscillator, translation metric, and Hermitian representative of Ref.~\cite{Dutta2013}, we state the bound-state domain explicitly and establish the reality of the point spectrum directly by endpoint analysis on the complex ray.}}  Second, we derive compact energy formulas and the leading $1/k$ corrections in both DSR prescriptions.  Third, we show that DSR changes the relative spinor components even when the spatial eigenfunctions remain unchanged.  Fourth, we analyze the AC critical point branch by branch and derive the corresponding Morse excitation threshold.  Finally, two figures display the pseudo-Hermitian Morse geometry and the qualitatively different MS/AC reconstruction maps.

The paper is organized as follows.  Section~II develops the GDO factorization and reviews the non-Hermitian consistency conditions needed below.  Section~III derives the MS and AC reconstruction maps, their spinor amplitudes, and their weak-deformation limits.  Section~IV treats the pseudo-Hermitian complex Morse interaction in detail, including shape invariance, eigenfunctions, DSR energies, and the excitation cutoff.  Section~V discusses the physical interpretation and limitations of the construction.  Section~VI summarizes the main conclusions.

\section{Generalized Dirac oscillator and non-Hermitian factorization}

\subsection{Dirac factorization in $(1+1)$ dimensions}\label{sec:factorization}

We set $c=1$ and keep $\hbar$ explicit.  Choosing \rev{$\chg{\alpha_{\rm D}}=\sigma_x$} and $\beta=\sigma_z$, the generalized Dirac oscillator is defined by
\begin{equation}
 p\longrightarrow p-i\beta f(x),
\end{equation}
with $p=-i\hbar\,d/dx$.  The Hamiltonian is
\begin{equation}
H_{\rm GDO}=\sigma_x p-\sigma_y f(x)+\sigma_z m.
\label{eq:Hgdo}
\end{equation}
For later use it is convenient to introduce the two first-order differential operators
\begin{equation}
\mathcal A=p-if(x),\qquad
\mathcal B=p+if(x).
\label{eq:AB}
\end{equation}
If $f$ is real, $\mathcal B=\mathcal A^\dagger$.  For complex $f$, this equality no longer holds in the standard inner product; below we show that $\mathcal B$ becomes the pseudo-adjoint of $\mathcal A$ for the Morse metric used in this work.

Writing $\Psi=(\psi_1,\psi_2)^T$, the stationary Dirac equation gives
\begin{align}
\mathcal B\psi_2&=(E-m)\psi_1,
\label{eq:coupled1}\\
\mathcal A\psi_1&=(E+m)\psi_2.
\label{eq:coupled2}
\end{align}
Elimination of either component yields
\begin{align}
H_-\psi_1\equiv\mathcal B\mathcal A\psi_1
&=\left[-\hbar^2\frac{d^2}{dx^2}+V_-(x)\right]\psi_1
=\epsilon\psi_1,
\label{eq:Hminus}\\
H_+\psi_2\equiv\mathcal A\mathcal B\psi_2
&=\left[-\hbar^2\frac{d^2}{dx^2}+V_+(x)\right]\psi_2
=\epsilon\psi_2,
\label{eq:Hplus}
\end{align}
where
\begin{equation}
V_\pm(x)=f^2(x)\pm\hbar f'(x),
\label{eq:partner}
\end{equation}
and, in the undeformed Dirac theory,
\begin{equation}
\epsilon=E^2-m^2.
\label{eq:undeformed_dispersion}
\end{equation}
Equations~\eqref{eq:Hminus}--\eqref{eq:partner} are the SUSY-QM factorization associated with the superpotential $f(x)$ \cite{Cooper1995,Junker2020}.  If a positive metric exists such that $\mathcal B=\mathcal A^\#$, the standard intertwining relations carry over with the pseudo-adjoint
\begin{equation}
\mathcal A^\#\equiv\eta^{-1}\mathcal A^\dagger\eta,
\label{eq:pseudoadjoint_def}
\end{equation}
\rev{where $\eta$ is an invertible Hermitian operator (the metric operator; for a physical inner product it is required to be positive).}
The partner operators are then isospectral for $\epsilon>0$, up to the usual possible zero mode of $H_-$.

An important distinction will be used later.  The normalized SUSY partner eigenfunction can be written as
\begin{equation}
\varphi_{+,n}=\frac{1}{\sqrt{\epsilon_n}}\mathcal A\varphi_{-,n},
\qquad \epsilon_n>0,
\label{eq:susy_partner_map}
\end{equation}
but the physical lower component of the Dirac spinor is not generally given by Eq.~\eqref{eq:susy_partner_map}.  In the undeformed theory, Eq.~\eqref{eq:coupled2} instead gives
\begin{equation}
\psi_{2,n}=\frac{1}{E_n+m}\mathcal A\psi_{1,n}.
\label{eq:dirac_lower_undeformed}
\end{equation}
The corresponding DSR prefactors will be derived in Sec.~III.

\subsection{Pseudo-Hermiticity and the role of $\mathcal{PT}$ symmetry}

A Hamiltonian $H$ is $\eta$-pseudo-Hermitian if
\begin{equation}
H^\dagger=\eta H\eta^{-1},
\label{eq:pseudoHerm}
\end{equation}
\rev{with $\eta$ the invertible Hermitian operator introduced after Eq.~\eqref{eq:pseudoadjoint_def}.}  If $\eta$ can be chosen positive and the spectral problem is diagonalizable with a complete set of states, the physical inner product may be written as
\begin{equation}
(\Phi,\Psi)_\eta=\langle\Phi|\eta|\Psi\rangle,
\label{eq:eta_product}
\end{equation}
and the theory is equivalent, through a similarity transformation, to a Hermitian representation \cite{Mostafazadeh2002,Mostafazadeh2010}.  \rev{For unbounded metrics, these statements require appropriate operator domains \cite{Mostafazadeh2013Unbounded}.  Here the identities and invertibility are used on the analytic bound-state eigenfunctions and their specified translated domains.  Questions of bounded similarity on the full reference space and completeness of scattering states lie outside the present bound-state construction.}

For comparison, conventional $\mathcal{PT}$ symmetry of a local Schr\"odinger operator requires
\begin{equation}
V(x)=V^*(-x).
\label{eq:PTcondition}
\end{equation}
When the $\mathcal{PT}$ symmetry is unbroken, the spectrum can be entirely real; spontaneous breaking produces complex-conjugate pairs \cite{Bender1998,Bender2007}.  In relativistic spinor systems the parity operation also acts on the spinor indices, so the precise Dirac-level condition is representation dependent.  For this reason, and because the Morse exponential is not parity even, our solvable example below is formulated directly in the pseudo-Hermitian framework.  The broader GDO construction can of course accommodate genuinely $\mathcal{PT}$-symmetric choices of $f(x)$, but that is not required for the spectral reality of the complex Morse model considered here.

\section{DSR reconstruction of the GDO spectrum}

The DSR realizations used here are the same single-particle wave-equation implementations employed in recent oscillator studies \cite{JafariBoumali2025PLB}.  \rev{We take $k>0$ and $m\ge0$; the particle/antiparticle branch interpretation below is used for $m<k$.  The closed-form roots solve the adopted first-order DSR wave equations exactly, while quantitative physical use is restricted to their low-energy regime $|E|/k\ll1$.}  The crucial feature for the present construction is that the differential operator $\mathcal B\mathcal A$ remains the same: DSR changes the algebraic relation between its eigenvalue $\epsilon_n$ and the physical energy, together with the relative amplitude of the two Dirac components.

\subsection{Magueijo--Smolin realization}

In the MS realization the mass term is replaced by the energy-dependent quantity
\begin{equation}
M(E)=m\left(1-\frac{E}{k}\right).
\label{eq:Meff}
\end{equation}
The coupled equations have the same form as Eqs.~\eqref{eq:coupled1}--\eqref{eq:coupled2} with $m$ replaced by $M(E)$.  Consequently,
\begin{equation}
\epsilon_n=E^2-m^2\left(1-\frac{E}{k}\right)^2.
\label{eq:MS_relation}
\end{equation}
For $m\neq k$ this gives the compact expression
\begin{equation}
E_{n,\sigma}^{\rm MS}=
\frac{-m^2/k+\sigma\sqrt{m^2+\epsilon_n\left(1-m^2/k^2\right)}}{1-m^2/k^2},
\qquad \sigma=\pm1.
\label{eq:MS_E_compact}
\end{equation}
The apparent denominator singularity at $m=k$ is not a singularity of the original quadratic equation: at that isolated parameter value Eq.~\eqref{eq:MS_relation} becomes linear in $E$.  In the physically relevant DSR regime $m/k\ll1$, Eq.~\eqref{eq:MS_E_compact} is regular.

The lower spinor component is reconstructed as
\begin{equation}
\psi_{2,n}^{\rm MS}=
\frac{1}{E_n+M(E_n)}\,\mathcal A\psi_{1,n}.
\label{eq:MS_spinor}
\end{equation}
Thus the spatial wavefunction $\psi_{1,n}$ is inherited from the undeformed factorized problem, while the relative spinor weight is DSR dependent.

\subsection{\chg{Amelino-Camelia} realization}

In the AC realization used in Ref.~\cite{JafariBoumali2025PLB}, the spatial Dirac operator is rescaled by
\begin{equation}
\kappa(E)=1+\frac{E}{2k}.
\label{eq:kappa}
\end{equation}
The separation relation is
\begin{equation}
\epsilon_n=\frac{E^2-m^2}{\left(1+E/2k\right)^2},
\label{eq:AC_relation}
\end{equation}
which gives
\begin{equation}
E_{n,\sigma}^{\rm AC}=
\frac{\epsilon_n/(2k)+\sigma\sqrt{m^2+\epsilon_n\left(1-m^2/4k^2\right)}}
{1-\epsilon_n/(4k^2)},
\qquad \sigma=\pm1.
\label{eq:AC_E_compact}
\end{equation}
The lower component is
\begin{equation}
\psi_{2,n}^{\rm AC}=
\frac{1+E_n/(2k)}{E_n+m}\,\mathcal A\psi_{1,n}.
\label{eq:AC_spinor}
\end{equation}
Equations~\eqref{eq:MS_spinor} and \eqref{eq:AC_spinor} show that the statement ``DSR acts only on the energy'' must be qualified: the spatial eigenvalue problem is unchanged in these realizations, but the physical spinor composition is also deformed.

\subsection{Undeformed limit and weak-deformation expansion}

Define
\begin{equation}
\Omega_n=\sqrt{m^2+\epsilon_n}.
\end{equation}
Both DSR prescriptions recover
\begin{equation}
E_{n,\sigma}\longrightarrow \sigma\Omega_n,
\qquad k\to\infty.
\end{equation}
Expanding the exact maps for large $k$ gives the particularly transparent results
\begin{align}
E_{n,\sigma}^{\rm MS}
&=\sigma\Omega_n-\frac{m^2}{k}+O(k^{-2}),
\label{eq:MS_expansion}\\
E_{n,\sigma}^{\rm AC}
&=\sigma\Omega_n+\frac{\epsilon_n}{2k}+O(k^{-2}).
\label{eq:AC_expansion}
\end{align}
Thus, at first order, both positive- and negative-energy branches are translated by the same amount, but in opposite directions for the two DSR realizations: the MS center shifts downward by $m^2/k$, whereas the AC center shifts upward by $\epsilon_n/(2k)$.  This immediately explains the loss of exact particle--antiparticle antisymmetry at finite $k$.

\begin{revblock}
A complementary nonrelativistic expansion makes the physical effect on a Morse ladder especially transparent.  Expanding the positive branch in $\epsilon_n/m^2\ll1$, while keeping the ratio $m/k$ unexpanded, gives
\begin{align}
E_{n,+}^{\rm MS}
&=\frac{m}{1+m/k}+\frac{\epsilon_n}{2m}
+O\!\left(\frac{\epsilon_n^2}{m^3}\right),\label{eq:NR_MS}\\
E_{n,+}^{\rm AC}
&=m+\left(1+\frac{m}{2k}\right)^2\frac{\epsilon_n}{2m}
+O\!\left(\frac{\epsilon_n^2}{m^3}\right).
\label{eq:NR_dsr}
\end{align}
Thus the MS realization shifts the positive rest level to $m/(1+m/k)$ while leaving the leading nonrelativistic excitation spacings unchanged.  The AC realization uniformly rescales those spacings by $(1+m/2k)^2$.  This distinction supplies a direct low-energy interpretation of the two reconstruction maps.
\end{revblock}

For reference, the main algebraic differences are summarized in Table~\ref{tab:maps}.

\begin{table}[t]
\caption{Spatial separation relation, lower-component reconstruction, and first-order energy shift for the undeformed, MS, and AC theories.  Here $\mathcal A=p-if(x)$ and $\Omega_n=\sqrt{m^2+\epsilon_n}$.}
\label{tab:maps}
\begin{ruledtabular}
\begin{tabular}{lccc}
Framework & Separation relation & $\psi_{2,n}/(\mathcal A\psi_{1,n})$ & $E_{n,\sigma}$ to $O(k^{-1})$\\
\hline
Undeformed & $E^2-m^2=\epsilon_n$ & $(E+m)^{-1}$ & $\sigma\Omega_n$\\
MS & $E^2-m^2(1-E/k)^2=\epsilon_n$ & $[E+M(E)]^{-1}$ & $\sigma\Omega_n-m^2/k$\\
AC & $(E^2-m^2)/(1+E/2k)^2=\epsilon_n$ & $(1+E/2k)/(E+m)$ & $\sigma\Omega_n+\epsilon_n/(2k)$\\
\end{tabular}
\end{ruledtabular}
\end{table}

\subsection{Generalized-eigenvalue weights and AC branch criticality}\label{sec:weights}

\begin{revblock}
The two adopted DSR equations can be written as generalized eigenvalue problems with an energy-independent weight.  Defining
\begin{equation}
K\equiv\sigma_xp-\sigma_y f(x),\qquad H_0\equiv K+\sigma_zm,
\end{equation}
one finds
\begin{align}
H_0\Psi&=E\,W_{\rm MS}\Psi,
& W_{\rm MS}&=I+\frac{m}{k}\sigma_z,\label{eq:MS_weight}\\
H_0\Psi&=E\,W_{\rm AC}\Psi,
& W_{\rm AC}&=I-\frac{K}{2k}.\label{eq:AC_weight}
\end{align}
On the $n$th SUSY doublet, with basis $(\varphi_{-,n},\varphi_{+,n})^T$ and $\epsilon_n>0$, Eq.~\eqref{eq:susy_partner_map} gives $K=\sqrt{\epsilon_n}\,\sigma_x$.  Hence
\begin{equation}
\operatorname{spec}W_{\rm MS}=\left\{1\pm\frac{m}{k}\right\},\qquad
\operatorname{spec}W_{\rm AC}=\left\{1\pm\frac{\sqrt{\epsilon_n}}{2k}\right\}.
\label{eq:weight_spectra}
\end{equation}
The MS weight is positive for $m<k$, while the AC weight on the $n$th doublet is positive precisely when
\begin{equation}
\epsilon_n<4k^2.
\label{eq:AC_condition}
\end{equation}
After the Morse similarity transformation, $H_0$ and these weights are Hermitian on the analytic bound-state subspace; equivalently, in the original representation the corresponding stationary modes are orthogonal in the weighted form $\langle\Psi_i|\eta W|\Psi_j\rangle$ for distinct real eigenvalues.
\end{revblock}

Equation~\eqref{eq:AC_E_compact} has a denominator that vanishes at
\begin{equation}
\epsilon_n=4k^2.
\label{eq:ACcritical}
\end{equation}
The two algebraic branches behave differently at this point.  As $\epsilon_n\to4k^2$ from below,
\begin{equation}
E_{n,+}^{\rm AC}\to +\infty,
\label{eq:ACplusdiv}
\end{equation}
whereas the negative branch has the finite removable limit
\begin{equation}
\lim_{\epsilon_n\to4k^2}E_{n,-}^{\rm AC}
=-k-\frac{m^2}{4k}.
\label{eq:ACminuslimit}
\end{equation}
\begin{revblock}
The weight formulation gives the critical value a sharper algebraic meaning.  For $\epsilon_n<4k^2$, the AC quadratic has negative root product and therefore one positive and one negative energy.  For $\epsilon_n>4k^2$ (with $m<2k$, which includes the branch regime used here), its discriminant remains non-negative, while
\begin{equation}
E_{n,+}^{\rm AC}+E_{n,-}^{\rm AC}
=\frac{\epsilon_n/k}{1-\epsilon_n/(4k^2)}<0,
\qquad
E_{n,+}^{\rm AC}E_{n,-}^{\rm AC}
=-\frac{m^2+\epsilon_n}{1-\epsilon_n/(4k^2)}>0.
\label{eq:AC_vieta}
\end{equation}
Both roots are then negative.  Thus Eq.~\eqref{eq:AC_condition} is exactly the positivity domain of the AC generalized-eigenvalue weight on the $n$th doublet and exactly the domain containing a positive-energy particle root of the adopted equation.  The separate low-energy requirement $|E|/k\ll1$ controls the range in which the truncated DSR wave equation can be used quantitatively; the immediate neighborhood of Eq.~\eqref{eq:ACcritical} lies outside that regime.
\end{revblock}

\subsection{Massless limit}

For $m=0$, the MS relation becomes exactly undeformed:
\begin{equation}
E_{n,\sigma}^{\rm MS}=\sigma\sqrt{\epsilon_n}.
\label{eq:MS_massless}
\end{equation}
The AC model remains deformed.  Writing $\omega_n=\sqrt{\epsilon_n}$,
\begin{equation}
E_{n,+}^{\rm AC}=\frac{2k\omega_n}{2k-\omega_n},
\qquad
E_{n,-}^{\rm AC}=-\frac{2k\omega_n}{2k+\omega_n}.
\label{eq:AC_massless}
\end{equation}
The positive branch exists for $\omega_n<2k$, consistently with the generalized-eigenvalue condition derived above.

\section{Pseudo-Hermitian complex Morse interaction}

\subsection{Metric operator and Hermitian representative}

We now choose
\begin{equation}
f(x)=D-qe^{-\alpha x},
\qquad
q=A+iB=|q|e^{i\phi},
\qquad D,\alpha>0.
\label{eq:morse_f}
\end{equation}
The corresponding partner potentials are
\begin{align}
V_-(x;D)
&=D^2+q^2e^{-2\alpha x}-q(2D+a)e^{-\alpha x},
\label{eq:morse_Vminus}\\
V_+(x;D)
&=D^2+q^2e^{-2\alpha x}-q(2D-a)e^{-\alpha x},
\label{eq:morse_Vplus}
\end{align}
where
\begin{equation}
a\equiv\hbar\alpha.
\label{eq:morse_a}
\end{equation}

\begin{figure}[H]
\centering
\includegraphics[width=0.98\linewidth]{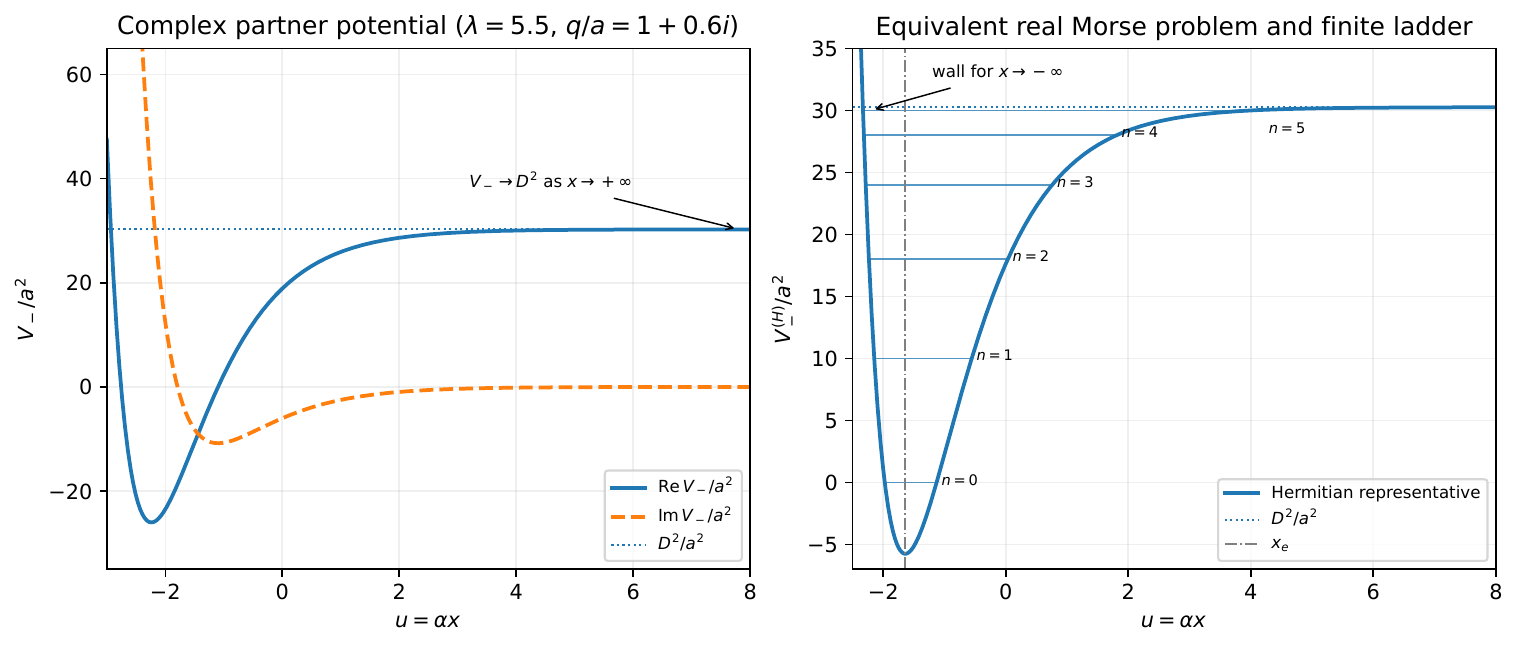}
\caption{Pseudo-Hermitian complex Morse interaction.  Left: real and imaginary parts of $V_-/a^2$ for the illustrative dimensionless choice $\lambda=D/a=5.5$ and $q/a=1+0.6i$.  Right: the Hermitian representative obtained by the half-shift similarity transformation, with its finite set of separation levels $\epsilon_n/a^2=2\lambda n-n^2$.  The horizontal level lines denote spectral eigenvalues; classical turning points are shown only for the Hermitian representative.  \rev{The coordinate ranges over the whole line $x\in\mathbb{R}$; the plotted window shows the approach $V_-\to D^2$ (continuum threshold) for $x\to+\infty$ and the \chg{rise toward} the wall for $x\to-\infty$, which make the potential reflection asymmetric.  \chg{The left panel extends to $\alpha x=-3$ so that the rising real part is visible; the right panel displays the Hermitian wall.}  In the right panel each level is drawn between its classical turning points, and the dash-dotted line marks the position $x_e$ of the minimum in Eq.~\eqref{eq:xe}.  For the displayed parameters, $\alpha x_e\simeq-1.6380$ and $\min V_-^{(H)}/a^2=-5.75$; the lower vertical limit is $-7$, so the well minimum is visible.}}
\label{fig:morse_metric}
\par\smallskip
{\small\rev{\chg{\textbf{Alt text:} Two panels show an asymmetric complex Morse potential and its real Hermitian representative versus $\alpha x$.  In the left panel the real part rises on the far left and approaches $30.25$ on the right; the imaginary part tends to zero on the right.  In the right panel the minimum is $-5.75$ at $\alpha x=-1.6380$, and six horizontal bound levels, $0$, $10$, $18$, $24$, $28$, and $30$, lie below the $30.25$ threshold.}}\par}
\end{figure}

\begin{revblock}
\emph{Range of $x$ and absence of reflection invariance.}\label{par:domain}  In contrast with the historical Morse potential, whose argument is the positive internuclear distance $r$ \cite{Morse1929}, the argument of Eq.~\eqref{eq:morse_f} is the one-dimensional coordinate of the $(1+1)$-dimensional Dirac equation, and the reference Hilbert space is $L^2(\mathbb{R},dx)\otimes\mathbb{C}^2$ with $-\infty<x<\infty$.  This is the range displayed in Fig.~\ref{fig:morse_metric}.  The two ends of the line are physically inequivalent.  For $x\to+\infty$ the superpotential tends to the constant $D$ and $V_\pm\to D^2$.  In the Hermitian representative, $\epsilon=D^2$ is the continuum threshold for scattering states entering from the right; our metric argument below is restricted to the discrete states.  For $x\to-\infty$ the term $q^2e^{-2\alpha x}$ dominates and, in the Hermitian representative constructed below, produces an infinitely high wall.  Hence the potential is not reflection invariant.  More precisely, with the Dirac parity operator $\mathcal P=\sigma_z\mathcal P_0$, where $\mathcal P_0\psi(x)=\psi(-x)$, Eq.~\eqref{eq:Hgdo} transforms as
\begin{equation}
\mathcal P\,H_{\rm GDO}[f(x)]\,\mathcal P^{-1}=H_{\rm GDO}[-f(-x)],
\label{eq:parity_GDO}
\end{equation}
so a GDO is parity invariant only when $f$ is odd, as for the linear Dirac oscillator $f\propto x$.  The Hermitian Morse function $f_H$ tends to $D$ for $x\to+\infty$ and to $-\infty$ for $x\to-\infty$; neither it nor the complex function is odd about any point.  Parity is therefore not a symmetry and the eigenstates carry no parity quantum number.  Figure~\ref{fig:morse_metric} displays the asymmetric potentials and their separation levels, not probability densities.  The reflected model with $\tilde f(x)=-f(-x)=-D+qe^{\alpha x}$ is unitarily equivalent by Eq.~\eqref{eq:parity_GDO} and therefore contains no new physics.  If one wishes to connect with the radial problem, $x$ can be identified with $r$ up to a constant shift, $x-x_e=r-r_e$, where $x_e$ is the minimum of the well [Eq.~\eqref{eq:xe} below] and $r_e$ the equilibrium distance.  A radial Morse problem with $r>0$ also requires a boundary condition at $r=0$ and is a different boundary-value problem.  Extending that half-line problem to the full line can be accurate when the boundary lies sufficiently far inside the forbidden region, where the full-line bound states are strongly suppressed [cf. Eq.~\eqref{eq:asymptotics}].  Its quantitative accuracy is therefore state-, parameter-, and boundary-condition dependent.  The present one-dimensional model takes the full line as its exact configuration space.\label{par:domain_end}

\emph{Parameters and their physical scales.}\label{par:scales}  With $c=1$, the superpotential $f$ has the dimension of momentum (energy), so $D$, $|q|$, and $a=\hbar\alpha$ are energies, $\alpha$ is an inverse length, and $\lambda\equiv D/a$ [Eq.~\eqref{eq:lambda_def}] is dimensionless.  The modulus $|q|$ is not a spectral parameter: a real translation $x\to x+\alpha^{-1}\ln(|q|/|q'|)$ changes $|q|$ into any other positive value, and it only fixes the position of the minimum of the Hermitian well,
\begin{equation}
x_e=\frac{1}{\alpha}\ln\frac{2|q|}{2D+a},
\qquad
\min_x V_-^{(H)}=D^2-\left(D+\frac{a}{2}\right)^2,
\label{eq:xe}
\end{equation}
while its phase $\phi$ is removed by the imaginary translation constructed below.  The separation spectrum therefore depends on $D$ and $a$ alone, and the full DSR problem on the four energy scales $D$, $a$, $m$, and $k$, i.e. on the dimensionless ratios $\lambda=D/a$ (number of bound states), $D/m$ (strength of relativistic effects, since $\epsilon_n\lesssim D^2$), and $m/k$ and $D/(2k)$ (strength of the DSR deformation; the latter controls the classification in Sec.~IV\,D).  The nonrelativistic limit fixes the correspondence with the historical Morse parameters: writing $E=m+E_{\rm NR}$ with $E_{\rm NR}\ll m$, one has $\epsilon\simeq2mE_{\rm NR}$, and Eq.~\eqref{eq:Hminus} becomes a Schr\"odinger equation with a Morse well of depth $D_e=(D+a/2)^2/(2m)$ and range parameter $\alpha$.  As an illustrative molecular-scale estimate, take $D_e=5\,$eV, $\alpha=2\,\text{\AA}^{-1}$, and interpret $m$ as a reduced mass in the range $0.5$--$10\,$GeV.  Then $a\simeq3.95\,$keV and $D=\sqrt{2mD_e}-a/2\simeq6.87\times10^4$--$3.14\times10^5\,$eV, giving $\lambda\simeq17.4$--$79.6$ and $D/m\ll1$.  These values illustrate the scale hierarchy rather than fit a particular molecule.  The historical Morse problem thus lies deep in the nonrelativistic and low-energy regime, whereas in MS/AC DSR the scale $k$ is of the order of the Planck energy, $k\sim E_{\rm P}\simeq1.2\times10^{28}\,$eV.  In that fundamental setting $D/(2k)\sim10^{-23}$ and only the case $D\le2k$ of Sec.~IV\,D is realized.  \chg{Equations~\eqref{eq:NR_MS}--\eqref{eq:NR_dsr} also show} the size of the observable low-energy effect: at leading nonrelativistic order the MS deformation changes the rest-energy offset but leaves Morse excitation spacings unchanged, whereas AC multiplies all spacings by $(1+m/2k)^2$.  For the illustrative reduced-mass range above, the AC fractional change is approximately $m/k\sim4\times10^{-20}$--$8\times10^{-19}$.  Moreover, a common rescaling $D\to cD$ and $\alpha\to c\alpha$ leaves $\lambda=D/(\hbar\alpha)$ unchanged while sending $\epsilon_n\to c^2\epsilon_n$; a uniform AC rescaling of one isolated ladder can therefore be absorbed into a refit of these Morse scales unless independent information fixes them.  The physical purpose of the model is accordingly twofold: (i) to provide an exactly solvable relativistic, anharmonic, and finite-depth generalization of the Dirac oscillator in which the interplay of non-Hermiticity, supersymmetry, and DSR kinematics can be followed analytically; and (ii) to provide analytic benchmarks for possible future analogue implementations inspired by trapped-ion and photonic Dirac-oscillator models \cite{Bermudez2007,Longhi2010,Sadurni2011}.  The cited analogue \chg{proposals} motivate this second direction.  A complete complex-Morse DSR implementation would additionally have to specify how an effective deformation scale and the phase of $q$ are engineered.  The case $D>2k$ is therefore retained here as a property of the algebraic AC map, with an effective-scale analogue interpretation remaining prospective.\label{par:scales_end}
\end{revblock}

For $B\neq0$ these potentials are non-Hermitian on the real axis.  \rev{\chg{Nevertheless, their point spectrum is real on the domain specified below, as proved directly in Sec.~IV\,B; the imaginary-translation construction supplies the corresponding Hermitian representation.}}  Let
\begin{equation}
\theta=\frac{2\phi}{\alpha},
\qquad
\eta=\exp\left(-\frac{\theta p}{\hbar}\right).
\label{eq:metric}
\end{equation}
Since
\begin{equation}
\eta x\eta^{-1}=x+i\theta,
\end{equation}
we have
\begin{equation}
f(x+i\theta)=D-qe^{-\alpha x}e^{-i\alpha\theta}
=D-q^*e^{-\alpha x}=f^*(x).
\label{eq:shiftcondition}
\end{equation}
It follows immediately that
\begin{equation}
H_{\rm GDO}^\dagger=\eta H_{\rm GDO}\eta^{-1}.
\label{eq:morse_pseudo}
\end{equation}
The same shift relation also identifies the factorization operator with the metric adjoint:
\begin{equation}
\mathcal A^\#=\eta^{-1}\mathcal A^\dagger\eta
=p+i f(x)=\mathcal B,
\label{eq:pseudoadjoint_morse}
\end{equation}
because $\eta^{-1}f^*(x)\eta=f^*(x-i\theta)=f(x)$.  Thus the SUSY factorization in Sec.~II is an exact pseudo-Hermitian factorization for this Morse family.

Furthermore, the positive square root
\begin{equation}
\rho=\eta^{1/2}=\exp\left(-\frac{\theta p}{2\hbar}\right)
\end{equation}
generates the half shift $x\to x+i\phi/\alpha$, under which
\begin{equation}
q e^{-\alpha(x+i\phi/\alpha)}=|q|e^{-\alpha x}.
\label{eq:halfshift}
\end{equation}
Hence
\begin{equation}
h=\rho H_{\rm GDO}\rho^{-1}
\label{eq:hermitian_rep}
\end{equation}
is the Hermitian GDO with the real Morse function
\begin{equation}
f_H(x)=D-|q|e^{-\alpha x}.
\label{eq:fH}
\end{equation}
\begin{revblock}
The translation argument is applied to the analytic bound states of Sec.~IV\,C.  The metric moves the real axis to $\operatorname{Im}x=\theta$.  Along the intermediate lines $\operatorname{Im}x=\tau$, with
\[
\min(0,\theta)\le\tau\le\max(0,\theta),
\]
the coefficient of $e^{-\alpha x}$ has phase $\phi-\alpha\tau$, which varies between $\phi$ and $-\phi$.  The wall-side factor in the magnitude of a bound state is $\exp[-(|q|/a)\cos(\phi-\alpha\tau)e^{-\alpha x}]$.  It decays throughout this interval in the adopted domain
\begin{equation}
A=\operatorname{Re}q>0,\qquad\text{i.e.}\qquad |\phi|<\frac{\pi}{2}.
\label{eq:admissible}
\end{equation}
Together with $D>0$ and the level condition derived in Sec.~IV\,B, this ensures square integrability of the displayed states and their images under $\rho$ and $\eta$.  For $\theta\ne0$, both $\eta$ and $\eta^{-1}$ are unbounded.  The similarity relations above are consequently used on these states and their specified translated domains, not as a bounded equivalence on the whole reference space \cite{Mostafazadeh2013Unbounded}.

Throughout this paper the finite positive-Morse ladder is constructed for $D>0$ and Eq.~\eqref{eq:admissible}.  Other sign domains require a separate endpoint analysis; the real-negative-$q$ case at fixed $D>0$ is discussed explicitly in Sec.~IV\,C.  The argument of $f_H$ in Eq.~\eqref{eq:fH} still ranges over the whole line $x\in\mathbb R$: the similarity transformation changes the representation of the states while preserving the configuration space.
\end{revblock}

\rev{For the admissible bound states, this explicit Hermitian representative makes two facts transparent: the separation eigenvalues are real, and the phase of $q$ changes the representation of the eigenfunctions but not these eigenvalues.}  \rev{\chg{This is the imaginary-shift construction for complex Morse potentials and generalized Dirac oscillators developed in Refs.~\cite{Ahmed2001,Dutta2013}.}}  It also shows why conventional $\mathcal{PT}$ symmetry is not needed for this particular solvable example.

Figure~\ref{fig:morse_metric} illustrates the complex potential and its Hermitian representative.  The horizontal lines in the right panel are the real separation eigenvalues derived below.

\subsection{Shape invariance and the finite spatial spectrum}\label{sec:spatial_spectrum}

The Morse pair in Eqs.~\eqref{eq:morse_Vminus}--\eqref{eq:morse_Vplus} is shape invariant.  Indeed,
\begin{equation}
V_+(x;D)=V_-(x;D-a)+R(D),
\qquad
R(D)=D^2-(D-a)^2.
\label{eq:shapeinv_morse}
\end{equation}
For unbroken SUSY, $H_-=\mathcal B\mathcal A$ has a zero mode \rev{(this requires $D>0$ and $\operatorname{Re}q>0$; see Sec.~IV\,C)} and the spectrum follows algebraically:
\begin{align}
\epsilon_n
&=\sum_{j=0}^{n-1}R(D-ja)
\nonumber\\
&=D^2-(D-na)^2
=2Dan-a^2n^2.
\label{eq:eps_morse}
\end{align}
Introducing
\begin{equation}
\lambda\equiv\frac{D}{a}=\frac{D}{\hbar\alpha},
\label{eq:lambda_def}
\end{equation}
and the dimensionless exponential coordinate
\begin{equation}
z\equiv\frac{2q}{a}e^{-\alpha x},
\label{eq:morse_z}
\end{equation}
\rev{we now state precisely which integers $n$ are allowed.  For real $x$, $z$ runs along the ray $\arg z=\phi$; in the Hermitian representative one replaces $q$ by $|q|$, so $z\in(0,\infty)$.}

\begin{revblock}
\emph{Definition of normalizability.}\label{par:normalizability}  A solution $\psi_1$ of Eq.~\eqref{eq:Hminus} is called normalizable if it is square integrable on the whole configuration space,
\begin{equation}
\int_{-\infty}^{\infty}|\psi_1(x)|^2\,dx<\infty ,
\label{eq:norm_def}
\end{equation}
in the Hermitian representative, or, equivalently for complex $q$ satisfying Eq.~\eqref{eq:admissible}, if its $\eta$-norm $(\psi_1,\psi_1)_\eta$ in Eq.~\eqref{eq:eta_product} is finite and positive.  Only such solutions are normalizable bound-state eigenfunctions; the upper component then fixes the lower one through Eqs.~\eqref{eq:dirac_lower_undeformed}, \eqref{eq:MS_spinor}, or \eqref{eq:AC_spinor}.  With $dx=-dz/(\alpha z)$ and $z\in(0,\infty)$ in the Hermitian representative, Eq.~\eqref{eq:norm_def} reads $\alpha^{-1}\int_0^\infty|\psi_1|^2\,dz/z<\infty$, so that the two ends $x\to+\infty$ ($z\to0$) and $x\to-\infty$ ($z\to\infty$) must be controlled separately.

\emph{Derivation of the bound on $n$.}  \chg{In the Hermitian representative, bound-state eigenvalues are real and satisfy} $\epsilon<D^2$, since $V_-\to D^2$ for $x\to+\infty$ and $V_-\to+\infty$ for $x\to-\infty$.  \chg{The complex-ray proof below does not assume this reality in advance.}  We define
\begin{equation}
s\equiv\frac{\sqrt{D^2-\epsilon}}{a}>0 ,
\label{eq:s_def}
\end{equation}
with the positive root.  For $x\to+\infty$ the two independent solutions behave as $e^{\mp s\alpha x}\propto z^{\pm s}$, and condition \eqref{eq:norm_def} selects $z^{+s}$.  Setting $\psi_1=z^{s}e^{-z/2}w(z)$, Eq.~\eqref{eq:Hminus} becomes Kummer's equation
\begin{equation}
z w''+(2s+1-z)w'-(s-\lambda)w=0,
\label{eq:kummer}
\end{equation}
whose solution regular at $z=0$ is $w={}_1F_1(s-\lambda;2s+1;z)$.  Unless the series terminates, its large-$z$ asymptotics contains an $e^z$ term \cite[Eq.~13.7.2]{DLMF}, and $\psi_1$ grows like $e^{z/2}$ under the wall \chg{in the Hermitian representative.  Termination} requires $s-\lambda=-n$ with $n=0,1,2,\ldots$, i.e. $s=\lambda-n$, and $w\propto L_n^{(2s)}(z)$.  Since $s$ is positive by construction, normalizability requires
\begin{equation}
n<\lambda,
\qquad n=0,1,2,\ldots.
\label{eq:morse_n_condition}
\end{equation}
The energy relation alone cannot provide this information.  Indeed, $\epsilon_n<D^2$ is equivalent to $(\lambda-n)^2>0$ and allows both $n<\lambda$ and $n>\lambda$.  The endpoint condition first fixes the positive exponent $s$, and polynomial termination then gives $s=\lambda-n>0$.  

\chg{\emph{Direct complex-ray proof without a reality assumption.}\label{par:complex_epsilon}
For $\epsilon\in[D^2,\infty)$, the large-$x$ solutions are oscillatory, or constant/linear at threshold, and no nonzero solution is square integrable.  For $\epsilon\in\mathbb C\setminus[D^2,\infty)$, choose $s=\sqrt{D^2-\epsilon}/a$ with $\operatorname{Re}s>0$.  Square integrability at $x\to+\infty$ selects $\psi_1=z^s e^{-z/2}{}_1F_1(s-\lambda;2s+1;z)$.  Along the ray $\arg z=\phi$ with $|\phi|<\pi/2$, one has $\operatorname{Re}z\to+\infty$ at the wall.  The exponentially growing contribution then has coefficient proportional to $\Gamma(2s+1)/\Gamma(s-\lambda)$ \cite[Eq.~13.7.2]{DLMF}.  The numerator is finite and nonzero because $\operatorname{Re}s>0$, so this contribution vanishes only when $s-\lambda=-n$, $n\in\mathbb N_0$.  Thus $s=\lambda-n$ is real and positive, and every square-integrable point eigenvalue is real and given by Eq.~\eqref{eq:eps_morse}.  This direct ray argument establishes the real quantization condition without using the unbounded metric.  It makes no claim of bounded similarity or completeness of scattering states on the whole reference space.\label{par:complex_epsilon_end}}

\label{par:formal_labels}Formal continuation of $z^{\lambda-n}e^{-z/2}L_n^{(2\lambda-2n)}(z)$ to $n>\lambda$ produces no additional states.  Generically the negative endpoint power is not canceled and the function is not square integrable at $z=0$.  At exceptional parameter values, the identity
\[
L_n^{(-r)}(z)=(-z)^r\frac{(n-r)!}{n!}L_{n-r}^{(r)}(z),
\qquad r=1,\ldots,n,
\]
which follows from the finite-sum representation \cite[Eq.~18.5.12]{DLMF}, cancels that negative power.  With $r=2(n-\lambda)$, the result is proportional to the already-counted state with label $n'=n-r=2\lambda-n$, whenever $n'$ is a non-negative integer.  For example, at $\lambda=1$ and $n=2$, $L_2^{(-2)}(z)=z^2/2$ and the formal expression is $ze^{-z/2}/2$, a multiple of the $n=0$ state.  Thus exceptional normalizable expressions are duplicates, not extra levels.  At integer $\lambda=n$, one instead has $s=0$ and $\epsilon=D^2$; the solution tends to a nonzero constant at $x\to+\infty$ and is a nonnormalizable threshold state.

The same bound follows from SUSY.  The $j$th Hamiltonian of the shape-invariant hierarchy has the superpotential $f_j(x)=(D-ja)-qe^{-\alpha x}$, whose zero mode is normalizable, i.e. SUSY is unbroken at step $j$, if and only if $D-ja>0$ (Sec.~IV\,C).  Here $\operatorname{Re}q>0$ remains fixed throughout the hierarchy.  Once $D-ja\le0$, the $H_-$ zero mode loses square integrability while the $H_+$ candidate diverges at the wall, so the zero-mode construction terminates.  Within the specified hierarchy with $\operatorname{Re}q>0$, this reproduces Eq.~\eqref{eq:morse_n_condition}.
\end{revblock}
Thus the largest allowed $n$ is the largest integer strictly smaller than $\lambda$; equivalently the number of bound states is $N_{\rm M}=\lceil\lambda\rceil$.  This wording avoids the common ambiguity at integer values of $\lambda$.  The levels satisfy
\begin{equation}
\rev{\epsilon_0=0,\qquad 0<\epsilon_n<D^2\quad(1\le n<\lambda).}
\label{eq:morse_bound}
\end{equation}

\subsection{Closed-form eigenfunctions}\label{sec:eigenfunctions}

The zero mode $\mathcal A\psi_{1,0}=0$ is
\begin{equation}
\psi_{1,0}(x)=\rev{\widetilde{\mathcal N}_{1,0}}
\exp\left[-\frac{D}{\hbar}x-\frac{q}{\hbar\alpha}e^{-\alpha x}\right].
\label{eq:morse_ground}
\end{equation}
\begin{revblock}
Its modulus is
\begin{equation}
|\psi_{1,0}(x)|=|\widetilde{\mathcal N}_{1,0}|
\exp\left[-\frac{D}{\hbar}x-\frac{A}{a}e^{-\alpha x}\right],
\qquad A=\operatorname{Re}q ,
\label{eq:asymptotics}
\end{equation}
so it decays as $e^{-Dx/\hbar}$ for $x\to+\infty$ if $D>0$, and doubly exponentially for $x\to-\infty$ if $A>0$.  This $H_-$ zero mode is therefore normalizable in the domain $D>0$, $\operatorname{Re}q>0$ adopted here, and SUSY is unbroken.  In the variable $z$ of Eq.~\eqref{eq:morse_z}, $\psi_{1,0}\propto z^\lambda e^{-z/2}$.  For real $x$, $z$ lies on the ray $\arg z=\phi$ with $|z|\in(0,\infty)$; in the Hermitian representative, $z\in(0,\infty)$.  The negative-real endpoint $z\to-\infty$ is reached for real negative $q$, not in the adopted domain.

For fixed $D>0$ and real $q<0$, the $H_-$ zero mode is absent and
\[
V_-=D^2+|q|^2e^{-2\alpha x}+|q|(2D+a)e^{-\alpha x}>D^2,
\]
so the $H_-$ sector has no bound states.  At $\operatorname{Re}q=0$ the displayed $H_-$ zero mode also loses square integrability.  Other sign choices can interchange the zero-mode partner.  In particular, for real $D<0$ and $q<0$,
\[
\mathcal B\chi_0=0,\qquad
\chi_0\propto\exp\!\left[\frac{D}{\hbar}x+\frac{q}{a}e^{-\alpha x}\right]
\]
is normalizable, so the zero mode belongs to $H_+=\mathcal A\mathcal B$ instead \cite{Cooper1995}.  The finite spectrum analyzed below is the positive-$D$, positive-$\operatorname{Re}q$ realization in which the zero mode belongs to $H_-$.

In that realization, $\mathcal A\psi_{1,0}=0$ while $H_+=\mathcal A\mathcal B$ has no normalizable zero mode, so the Dirac zero mode is $\Psi_0=(\psi_{1,0},0)^T$.  For $m>0$, it belongs only to the positive-energy branch:
\[
E_0=m\quad\hbox{(undeformed and AC)},\qquad
E_0^{\rm MS}=\frac{m}{1+m/k}.
\]
Accordingly, the negative Dirac branch starts at $n=1$ when $m>0$.  At $m=0$, the isolated state instead has $E_0=0$ in all three frameworks; the coincident algebraic signs do not create two independent states.
\end{revblock}
\par

Using the variable $z$ defined in Eq.~\eqref{eq:morse_z}, \rev{the bound states, including $n=0$, are}
\begin{equation}
\psi_{1,n}(x)=\mathcal N_{1,n}
 z^{\lambda-n}e^{-z/2}L_n^{(2\lambda-2n)}(z),
\qquad n<\lambda.
\label{eq:morse_wavefunction}
\end{equation}
In the Hermitian representative one replaces $q$ by $|q|$, so that $z$ is real and positive.  \rev{The normalization constant for a unit spatial upper-component norm is then}
\begin{equation}
|\mathcal N_{1,n}|^2
=\alpha\,\frac{(2\lambda-2n)n!}{\Gamma(2\lambda-n+1)}.
\label{eq:morse_norm}
\end{equation}
\begin{revblock}
For complex $q$ in the adopted domain, the same modulus in Eq.~\eqref{eq:morse_norm} gives a unit spatial $\eta$-norm because $\rho\psi_{1,n}$ is the corresponding Hermitian eigenfunction.  The coefficient in Eq.~\eqref{eq:morse_ground} differs from the $n=0$ coefficient in Eq.~\eqref{eq:morse_wavefunction}: since $z^\lambda=(2q/a)^\lambda e^{-Dx/\hbar}$, the two conventions are related by
\[
\widetilde{\mathcal N}_{1,0}=\mathcal N_{1,0}(2q/a)^\lambda,
\qquad
|\widetilde{\mathcal N}_{1,0}|^2
=\frac{\alpha(2|q|/a)^{2\lambda}}{\Gamma(2\lambda)}.
\]
The power is taken continuously on $|\arg q|<\pi/2$.

Equation~\eqref{eq:morse_norm} normalizes the spatial upper component.  For $n\ge1$, let $\varphi_{-,n}$ and $\varphi_{+,n}=\mathcal A\varphi_{-,n}/\sqrt{\epsilon_n}$ have unit spatial $\eta$-norms.  The spinor is proportional to $(\varphi_{-,n},r_n\varphi_{+,n})^T$, where
\[
r_n^{(0)}=\frac{\sqrt{\epsilon_n}}{E_n+m},\qquad
r_n^{\rm MS}=\frac{\sqrt{\epsilon_n}}{E_n+M(E_n)},\qquad
r_n^{\rm AC}=\frac{\kappa(E_n)\sqrt{\epsilon_n}}{E_n+m}.
\]
For the deformed stationary equations the natural generalized-eigenvalue bilinear form is the weighted product $\langle\Psi|\eta W_{\rm MS,AC}|\Psi\rangle$ introduced in Eqs.~\eqref{eq:MS_weight}--\eqref{eq:AC_weight}, rather than the componentwise $\eta$ product alone.  It is positive on the MS bound-state sector for $m<k$ and on an AC $n$th doublet for $\epsilon_n<4k^2$.  The coefficient $r_n$ fixes the relative spinor amplitude, after which the overall constant may be chosen to give unit weighted norm.  This stationary normalization is the one used for the generalized eigenvalue problem.
\end{revblock}

\subsection{DSR-deformed Morse energies and excitation thresholds}\label{sec:thresholds}

Substitution of Eq.~\eqref{eq:eps_morse} into Eqs.~\eqref{eq:MS_E_compact} and \eqref{eq:AC_E_compact} \rev{gives the exact energy roots of the adopted DSR wave equations.}  For compactness, define
\begin{equation}
\epsilon_n=a^2n(2\lambda-n),
\qquad n<\lambda.
\end{equation}
Then
\begin{equation}
E_{n,\sigma}^{\rm MS}=
\frac{-m^2/k+\sigma\sqrt{m^2+a^2n(2\lambda-n)(1-m^2/k^2)}}{1-m^2/k^2},
\label{eq:MS_morse}
\end{equation}
and
\begin{equation}
E_{n,\sigma}^{\rm AC}=
\frac{a^2n(2\lambda-n)/(2k)
+\sigma\sqrt{m^2+a^2n(2\lambda-n)(1-m^2/4k^2)}}
{1-a^2n(2\lambda-n)/(4k^2)}.
\label{eq:AC_morse}
\end{equation}
\begin{revblock}
The signs in Eqs.~\eqref{eq:MS_morse}--\eqref{eq:AC_morse} must be combined with the Dirac zero-mode restriction.  For $0<m<k$, the physical index ranges are
\[
\begin{array}{ll}
\sigma=+1:&0\le n<\lambda,\\[2pt]
\sigma=-1:&1\le n<\lambda,
\end{array}
\]
with the additional condition $\epsilon_n<4k^2$ for the positive-energy AC particle root.  At $m=0$, the $n=0$ state is counted once at $E=0$, and the two signs label the nonzero modes.  Physical use of any of these roots further requires the low-energy condition $|E_{n,\sigma}|/k\ll1$; the index bounds alone do not ensure it.
\end{revblock}

The Morse ladder already terminates because $n<\lambda$.  The AC particle branch can terminate earlier.  \rev{Which of the two cases below applies depends on the ratio $D/(2k)$ discussed in Sec.~IV\,A.  If $k$ is the Planck energy and the Morse parameters take molecular or other low-energy values, $D/(2k)$ is negligibly small and only the first case occurs.  The second case is a statement about the mathematical structure of the AC map.  Its interpretation in a possible analogue system with an effective deformation scale $k_{\rm eff}$ would require a separate implementation.}  If $D\leq2k$, then $\epsilon_n<D^2\leq4k^2$, and every intrinsic Morse bound state lies below the AC critical value.  If $D>2k$, the additional condition $\epsilon_n<4k^2$ becomes
\begin{equation}
n<n_{\rm AC}^{\rm crit}
\equiv\frac{D-\sqrt{D^2-4k^2}}{a}
=\lambda-\sqrt{\lambda^2-\left(\frac{2k}{a}\right)^2}.
\label{eq:ncrit}
\end{equation}
Since $D>2k$ implies $n_{\rm AC}^{\rm crit}<\lambda$, the AC condition is then automatically the stricter one, and the positive-energy AC sector contains
\begin{equation}
0\leq n<n_{\rm AC}^{\rm crit}
\qquad(D>2k).
\label{eq:combined_cutoff}
\end{equation}
\rev{The bound in Eq.~\eqref{eq:combined_cutoff} is an exact property of the adopted AC generalized eigenvalue problem; quantitative use still requires the independent low-energy condition $|E|/k\ll1$.}

\begin{figure}[H]
\centering
\includegraphics[width=0.90\linewidth]{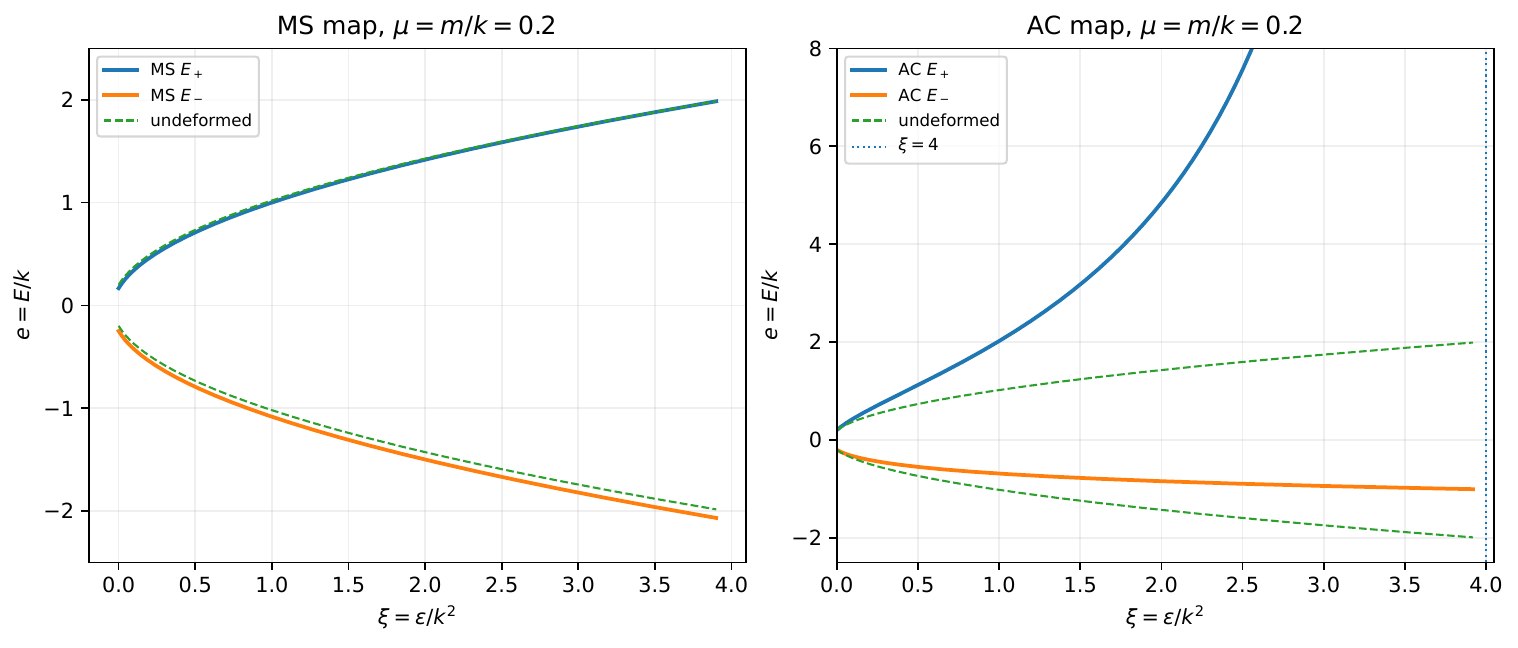}
\caption{Dimensionless DSR reconstruction maps for $\mu=m/k=0.2$.  Dashed curves show the undeformed result $e=\pm\sqrt{\mu^2+\xi}$.  In the MS model the two branches are displaced asymmetrically about zero.  In the AC model the positive branch diverges as $\xi\to4^-$, while the negative branch stays finite, approaching $-1-\mu^2/4$.  \rev{The plotted value of $\mu$ exaggerates the deformation to expose the algebraic structure; the near-critical curves are shown for that purpose, while quantitative use of the truncated wave equation is restricted to $|E|/k\ll1$.}}
\label{fig:dsr_maps}
\par\smallskip
{\small\rev{\chg{\textbf{Alt text:} Two panels compare MS and AC energy branches with dashed undeformed branches for $m/k=0.2$.  In MS, the branches are shifted asymmetrically about zero.  In AC, the positive branch rises steeply as $\epsilon/k^2$ approaches $4$, while the negative branch remains finite and tends to $-1.01$.  The dotted vertical line marks the algebraic boundary at $4$.}}\par}
\end{figure}

To display the DSR effect independently of any particular Morse parameter set, introduce the dimensionless quantities
\begin{equation}
\xi=\frac{\epsilon}{k^2},\qquad
\mu=\frac{m}{k},\qquad
e=\frac{E}{k}.
\end{equation}
Figure~\ref{fig:dsr_maps} compares the two reconstruction maps.  The moderate value $\mu=0.2$ is deliberately chosen to make the deformation visible and serves as an illustrative effective-scale plot.

\FloatBarrier

\section{Discussion}

\rev{The analysis separates three logically distinct structures.}  First, the complex interaction determines whether the spatial operator is mathematically consistent.  For the Morse family, the direct Kummer argument on $\arg z=\phi$ gives the real discrete quantization condition, while the metric in Eq.~\eqref{eq:metric} and the Hermitian representative in Eq.~\eqref{eq:hermitian_rep} provide the corresponding pseudo-Hermitian representation.  Second, shape invariance determines the finite set of separation constants $\{\epsilon_n\}$.  Third, the DSR prescription maps each allowed $\epsilon_n$ to a relativistic energy and changes the relative spinor amplitudes.  This layered formulation makes it straightforward to replace the Morse interaction by another solvable pseudo-Hermitian or $\mathcal{PT}$-symmetric GDO without rederiving the DSR kinematics.

The weak-deformation and nonrelativistic expansions, Eqs.~\eqref{eq:MS_expansion}--\eqref{eq:NR_dsr}, give a simple physical diagnostic.  At first order the MS correction is controlled by the rest-mass scale, whereas the AC correction is controlled directly by the spatial excitation $\epsilon_n$.  In the nonrelativistic particle sector the distinction is even sharper: MS shifts the rest offset while preserving the leading excitation spacings, whereas AC rescales the whole ladder.  The exact formulas also show that branch asymmetry follows from the non-even dependence on $E$ in the deformed dispersion relations.

\begin{revblock}
The scope of the present results is the stationary bound-state sector of the adopted first-order DSR wave equations.  The complex-Morse similarity transformation is used on the analytic bound-state domain specified in Sec.~IV, and the DSR equations are quantitatively controlled in the low-energy regime $|E|/k\ll1$.  The near-critical AC curves and the case $D>2k$ expose the exact algebraic structure of those adopted equations, while fundamental-Planck-scale molecular parameters satisfy $D/(2k)\sim10^{-23}$ and remain far from that boundary.  Analogue realizations with an effective deformation scale \chg{remain a prospective application, motivated by trapped-ion and photonic Dirac-oscillator proposals \cite{Bermudez2007,Longhi2010,Sadurni2011}; implementing the combined complex-Morse DSR system would require a separate construction.}
\end{revblock}

Finally, the massless limit provides a sharp model discriminator.  The MS deformation disappears identically when $m=0$ in the realization used here, while the AC map remains deformed because its modification acts in the momentum sector.  This contrast is independent of the detailed spatial interaction and is therefore a general feature of the reconstruction formulas.

\section{Conclusion}

We have developed a more explicit formulation of the generalized Dirac oscillator with a complex Morse interaction in Magueijo--Smolin and \chg{Amelino-Camelia} doubly special relativity.  The main improvement is the separation of the problem into well-defined mathematical layers.  \rev{On the adopted domain $D>0$, $\operatorname{Re}q>0$, the analytic bound states have a direct Kummer quantization on the ray $\arg z=\phi$ and an explicit positive translation metric leading to a Hermitian Morse representative.}  Shape invariance then produces the finite real spectrum $\epsilon_n=D^2-(D-n\hbar\alpha)^2$ with the unambiguous normalizability condition $n<D/(\hbar\alpha)$\rev{, obtained from square integrability on the whole line $x\in\mathbb{R}$.  In this realization $H_-$ carries the normalizable zero mode; the associated Dirac level belongs to the positive-energy branch for $m>0$ and becomes a single zero-energy state for $m=0$.}  The DSR prescriptions subsequently reconstruct the relativistic energies and the lower spinor amplitudes without modifying the spatial eigenfunctions in the realizations considered.

The compact energy formulas reveal distinct leading corrections: $-m^2/k$ in the MS model and $+\epsilon_n/(2k)$ in the AC model.  \rev{Their generalized-eigenvalue form supplies an exact algebraic criterion: $W_{\rm MS}$ is positive for $m<k$, while $W_{\rm AC}$ on the $n$th doublet is positive precisely for $\epsilon_n<4k^2$.  Across the latter boundary the AC root pair changes from one positive and one negative root to two negative roots, with the negative root retaining the finite limit in Eq.~\eqref{eq:ACminuslimit}.}  For $D>2k$, Eq.~\eqref{eq:ncrit} gives $n_{\rm AC}^{\rm crit}<\lambda$, so the AC condition is the stricter excitation bound.  \rev{For molecular scales at fundamental $k$, however, this boundary is remote; the leading nonrelativistic MS spacing is unchanged, whereas the AC spacing correction is only of relative order $m/k\sim10^{-19}$ for the illustrative masses considered here.}  In the massless limit the MS map reduces exactly to the undeformed dispersion relation, whereas the AC deformation survives.

These results turn the GDO into a modular testbed for comparing non-Hermitian spectral methods with deformed relativistic kinematics.  Natural extensions include genuinely $\mathcal{PT}$-symmetric GDO interactions, other shape-invariant complex potentials, and thermodynamic spectral sums in which the finite Morse ladder and the AC particle-branch truncation are treated consistently.  A further useful direction would be to formulate the same construction in DSR realizations beyond the low-energy wave equations used here, so that the behavior near the deformation scale can be assessed without relying on a truncated expansion.

\begin{acknowledgments}
\rev{\chg{The author thanks the referee for constructive comments that helped clarify the physical scales, coordinate domain, supersymmetry, and normalizability conditions.}}
\FundingStatement
\end{acknowledgments}

\end{document}